\documentclass[letterpaper,10 pt, conference]{ieeeconf}

\let\labelindent\relax
\usepackage{amsthm,amssymb,amsmath}
\usepackage{graphicx}
\usepackage{subfigure}
\usepackage{cite}
\usepackage{enumerate}
\usepackage{enumitem}
\usepackage{multirow, multicol}
\usepackage[english]{babel}
\usepackage{subfig}
\usepackage{tikz}
\usetikzlibrary{shapes,arrows,calc,positioning,arrows.meta}
\usepackage{nomencl}
\usepackage{cleveref}
\usepackage{upgreek}
\usepackage{mathrsfs}
\usepackage{upgreek}
\usepackage{mathtools, cuted}
\usepackage{flushend}
\usepackage{algorithm}
\usepackage{algpseudocode}

\newcommand{\beq}{\begin{equation}}
\newcommand{\eeq}{\end{equation}}
\newcommand{\benn}{\begin{equation*}}
\newcommand{\eenn}{\end{equation*}}

{}
{}
{}

\IEEEoverridecommandlockouts                              
\begin{document}
\tikzset{
block/.style = {draw, fill=white!20, rectangle, thin, minimum height=2.0em, minimum width=1.50em},
tmp/.style  = {coordinate}, 
sum/.style= {draw, fill=white!20, circle, node distance=.5cm},
tip/.style = {->, >=stealth', thin, dashed},
input/.style = {coordinate},
output/.style= {coordinate},
pinstyle/.style = {pin edge={to-,thin,black}},
startstop/.style = {draw, rounded rectangle, text centered, draw=black,thick},
io/.style = {trapezium, trapezium left angle=70, trapezium right angle=110, text centered},
process/.style = {rectangle, text centered, draw=black,thick},
decision/.style = {diamond, text centered, draw=black,thick},
arrow/.style = {-{Stealth[scale=1.2]},rounded corners,thick}
}
\title{Real-Time Cubature Kalman Filter Parameter Estimation of Blood Pressure Response Characteristics Under Vasoactive Drugs Administration*}
%
%
%

\author{
        Shahin~Tasoujian,
        Saeed~Salavati,
        Karolos~Grigoriadis,
        and~Matthew~Franchek
\thanks{* This paper is a preprint of a paper submitted to the IEEE  American Control Conference 2020.}
\thanks{ S. Tasoujian, S. Salavati, K. Grigoriadis, and M. Franchek are with the Department of Mechanical Engineering at the University of Houston, Houston, TX 77004 USA (e-mail: \tt\small  stasoujian, ssalavatidezfuli, karolos, mfranchek@uh.edu).}}
\maketitle

\begin{abstract}
Mathematical modeling and real-time dynamics identification of the mean arterial blood pressure (MAP) response of a patient to vasoactive drug infusion can provide a reliable tool for automated drug administration and therefore, reduce the emergency costs and significantly benefit the patient's MAP regulation in an intensive care unit. To this end, a dynamic first-order linear parameter-varying (LPV) model with varying parameters and varying input delay is considered to capture the MAP response dynamics. Such a model effectively addresses the complexity and the intra- and inter-patient variability of the physiological response. We discretize the model and augment the state vector with model parameters as unknown states of the system and a Bayesian-based multiple-model square root cubature Kalman filtering (MMSRCKF) approach is utilized to estimate the model time-varying parameters. Since, unlike the other model parameters, the input delay cannot be captured by a random-walk process, a multiple-model module with a posterior probability estimation is implemented to provide the delay identification. Validation results confirm the effectiveness of the proposed identification algorithm both in simulation scenarios and also using animal experiment data.
\end{abstract}


%
\IEEEpeerreviewmaketitle

\section{Introduction}
%
%
%
%
Fast-acting vasoactive medications are often used as a potentially vital medical intervention to address the patients' hemodynamics instability and regulate the blood pressure to a desired target value in numerous clinical and emergency resuscitation procedures. The vasoactive drugs are divided into two main categories: (1) vasodilator drugs that are being administered to treat individuals with hypertension. This type of drug widens blood vessels, and thereby, can rapidly lower the blood pressure. One of the most effective vasodilator drugs is sodium nitroprusside (SNP) which has been used to treat elevated blood pressure in various clinical scenarios such as post-surgical care, childbirth, and treating the typical high blood pressure disorders \cite{Craig2004, DaSilva2019}. (2) Vasopressor drugs such as phenylephrine (PHP), vasopressin, and norepinephrine increase the blood pressure by stimulating the depressed cardiovascular system causing vasoconstriction. These medicines are being used to treat patients with hypotensive symptoms in different medical scenarios such as hemorrhage, traumatic brain injury, and septic shock \cite{herget2008approach}.

Recently, the automated closed-loop administration of vasoactive drugs for the mean arterial pressure (MAP) control and regulation purposes, has gained significant attention in clinical care \cite{StasoujianRobust, DaSilva2019}. The automated dosage administration procedure utilizes the wealth of feedback control and surpasses the manual drug infusion with a syringe, in terms of accuracy, timeliness, cost efficiency, and reliability. Nevertheless, in order to have an accurate automated drug administration and to be able to implement robust model-based control strategies, an explicit dynamical model that accurately describes MAP response behavior to the drug infusion is needed. However, based on experiments, there are significant variations in the patient's physiological response to the drug infusion \cite{kashihara2004adaptive}. This pharmacological variation of the patient's MAP response to the drug causes the model parameters to vary over time for an individual, as well as, from patient-to-patient. Consequently, due to such intra- and inter-patient physiological variations, a mathematical model with fixed parameters cannot be adequate to capture the patient's complex MAP dynamics. Moreover, although a robust controller design can guarantee the stability and target MAP tracking of the patient's varying closed-loop system, other properties such as the settling time, rising time, damping characteristics, and disturbance rejection can significantly degrade when the patient’s parameters are off the control parameters.  In this regard, a real-time MAP response modeling and efficient parameter estimation scheme is essential.

There are multiple approaches for the estimation of dynamical system parameters. In the first approach, a linearization of the dynamical system is used to perform the parameter estimation. Extended Kalman filtering (EKF) is one of the widely used methods. However, it is only applicable to systems with mild nonlinearities, and it requires the Jacobian matrix computation. Moreover, numerical errors due to truncation and convergence problems are likely in EKF and other local approximation based estimators \cite{Arasaratnam2009}. In the other approach, known as the sampling method, the nonlinear representation of a system is used to estimate the parameters via a filter such as the unscented Kalman filter (UKF) which leads to more accurate estimation. In UKF, a set of weighted sampling points propagates via the nonlinear function of the system. However, for higher-order systems, UKF is prone to numerical instability since the weights of the sigma points may become negative \cite{Zhao2018}. Another sampling method is particle filtering (PF) which is an iterative Monte Carlo based method to compute the posterior probability distribution of the state of a nonlinear system even with non-Gaussian noise. PF requires a large set of randomly generated particles to approximate the posterior probability density function. Under an increase in the number of iterations, PF encounters particle degradation and depletion. In order to overcome such issues, the authors in \cite{Arasaratnam2009} have proposed a Bayesian filtering framework known as cubature Kalman filtering (CKF). The sample points in the CKF algorithm propagate via equally valued cubature points which are twice the size of the system nonlinear function. It uses a spherical-radial cubature rule to generate the weighted sum of sampling points to approximate the integrals in Bayesian estimation. CKF demonstrates better nonlinear performance, stability, and accuracy compared to EKF, PF, and UKF \cite{Zhao2018}.

In \cite{Furutani2004}, the authors have used a first-order model with delayed measurements to describe the MAP dynamics in response to hypotensive drugs. They have pre-identified the parameters using dose-response characteristics induced by a rectangular test signal while trying to avoid any adverse effect on the patient. If the identified parameters are not within the prescribed bounds, then the experiment will be repeated. Nonetheless, in the case of outlying identification results, the worst-case parameters are used. The output is filtered by a constant filter which has been derived using trial and error. The delay is determined via the response settling characteristics. 
In another work, a generalized fuzzy neural network framework has been studied for the estimation and control of MAP dynamics in response to vasodilator drugs \cite{Gao2005}. The parameters have been assumed to be nonlinear functions of the measured MAP. This method requires a training dataset and an effective learning algorithm for the artificial scheme. Moreover, overparameterization and a proper number of perceptrons remain as other obstacles. 
Based on the time delay model introduced in \cite{Slate1979}, the authors in \cite{Zhu2008} have proposed discrete-time parameters update laws. However, the procedure of the parameters identification of the original model has not been addressed.
A bank of Kalman filters (KFs) augmented with a posterior probability estimator to match a candidate model to that of the patient has been designed in \cite{Malagutti2013}. Each KF is responsible for generating the state vector updates for the next step, and the Kalman gain is assumed to be generated a priori. Then, by calculating the residual of the actual and generated output, the state vector is updated accordingly. In order to capture the varying time delay, which a conventional KF is not capable of, the multiple-model (MM) approach has been adopted through which five equally spaced delay blocks from $10s$ to $50s$, each is considered to be cascaded with the same bank of KFs. The recursive posterior probability estimation is calculated for each residual to update the input with the most likely delay. In a similar approach, \cite{DaSilva2019} has examined KF for the estimation of the MAP dynamics parameters in hypertension. In this work, authors have utilized the model introduced first in \cite{Slate1979}. They have discretized and transformed the infinite-dimensional model into a linear one that accommodates an input with three backward steps. Then, the parameters are gathered in a vector that is updated through the KF approach. The control parameters are updated using heuristic methods, a condition on the updated sensitivity along with curve fitting through polynomial regression. However, it should be noted that the conventional KF algorithm's convergence can only be guaranteed in an ideal linear-Gaussian environment. 
Reference \cite{Malagutti2014} has addressed the marginalized PF design in order to estimate the model parameters in the case of hypertension under SNP administration. The method allows considering linear and nonlinear states to be estimated separately to reduce the computational burden. 

The present work utilizes a multiple-model square root CKF (MMSRCKF) method to effectively estimate the nonlinear MAP model parameters in the case of hypotension under a vasopressor drug injection. The patient's MAP response model is described by a first-order parameter-varying model with a varying input delay. The multiple-model part addresses the hypothesis testing and the estimation of the input delay. The square root (SR) algorithm employs the Cholesky factorization of the error covariance matrix to guarantee its positive definiteness during numerical operations \cite{Cui2019}. For the verification of the proposed method, data from  animal experiments is collected at the University of Texas, Medical Branch (UTMB) at Galveston. The estimation results are compared to that of MMEKF reported in \cite{luspay2016adaptive}.

The mathematical notation to be used in the paper is as follows. $t$ denotes the continuous-time domain, and $k$ stands for the discrete-time variable. For a stochastic process, $\mathbf{x}_{k}$, $\mathscr{E}[\mathbf{x}_{k}]$ denotes its expected value and $\mathscr{N}\{\mathbf{x}_k;\widehat{\mathbf{x}}_{k|k},\mathbf{P}_{k|k} \}$ represents a normal Gaussian probability distribution with the mean of $\widehat{\mathbf{x}}_{k|k}$ and the covariance of $\mathbf{P}_{k|k}$.

This paper is organized as follows. The MMSRCKF online estimation algorithm is developed in Section \ref{sec:methodology}. Section \ref{sec:ProbForm} introduces a first-order time-delayed dynamic model to characterize the MAP response to the drug infusion. Section \ref{sec:EstimationResults} presents the estimation results and evaluation of the performance of the proposed MMSRCKF method in comparison to MMEKF. Final remarks are provided in Section \ref{sec:conclusion}.

\section{Estimation Preliminaries and Methodology}\label{sec:methodology}
In this section, first, a derivative-free on-line sequential state estimator known as the square root CKF (SRCKF) algorithm is formulated for a general nonlinear discrete-time stochastic system. Subsequently, we formulate the multiple-model (MM) approach and couple it with the introduced SRCKF algorithm for the time delay estimation of a system with an input delay.

\subsection{SRCKF Algorithm}
	The Bayesian-based CKF scheme aims at estimating the states of a dynamical system using a probabilistic framework \cite{Arasaratnam2009}. The original CKF state estimation process is susceptible to numerical problems such as indefinite error covariance matrix, divergence phenomenon, and filter instability.  In order to tackle these obstacles, we enhance CKF with the square root computation, \textit{i.e.} the covariance matrix is decomposed using a factorization method, such as the Cholesky factorization to guarantee positive definiteness within numerical operations \cite{loehr2014advanced}. The resulting square roots of the error covariance matrices propagate through the sequential state estimation process. Next, the third-degree spherical-radial rule is used to approximate the multidimensional integrals involved in the Bayesian filtering \cite{jia2013high}. 
Let us consider a general nonlinear discrete-time stochastic system as follows
	\beq
		\left\{\begin{array}{l}
		\mathbf{x}_{k+1} = \mathbf{f}(\mathbf{x}_k,\mathbf{u}_{k}) + \mathbf{w}_k,\\
		\mathbf{y}_k = \mathbf{h}(\mathbf{x}_k,\mathbf{u}_k) + \mathbf{v}_k,\; k=0,1,\ldots,k_f,
		\end{array}\right.
	\eeq
\noindent where $\mathbf{x}_k \in \mathbb{R}^n$ stands for the unmeasured state vector of the system, $\mathbf{u}_{k} \in \mathbb{R}^{n_u}$ is the input vector, and $\mathbf{y}_k \in \mathbb{R}^{n_y}$ is the measurement vector at the time $k$, and $k_f$ is the final time. $\mathbf{f}(\mathbf{x}_k,\mathbf{u}_k): (\mathbb{R}^n,\mathbb{R}^{n_u}) \mapsto \mathbb{R}^n$ and $\mathbf{h}(\mathbf{x}_k,\mathbf{u}_k): (\mathbb{R}^n,\mathbb{R}^{n_u}) \mapsto \mathbb{R}^{n_y}$ are known general nonlinear vector mappings, and $\mathbf{w}_k \in \mathbb{R}^n$ and $\mathbf{v}_k \in \mathbb{R}^{n_y}$ are statistically independent zero-mean Gaussian process and measurement noise signals, respectively. The probability distribution functions (PDFs) of the noise vectors, namely $p(\mathbf{w}_k)$ and $p(\mathbf{v}_k)$ are known, as well as, the initial state vector PDF $p(\mathbf{x}_0)$.

SRCKF seeks to find the estimation of the state vector in the form of a conditional PDF, $p(\mathbf{x}_k|\mathbf{y}^k)$, that has the entire knowledge about the current state vector, $\mathbf{x}_k$, given the entire measurement vectors sequence, \textit{i.e.} \textcolor{black}{$\mathbf{y}^k=[\begin{array}{cccc} \mathbf{y}_0 & \mathbf{y}_1 & \ldots & \mathbf{y}_k\end{array}]$.}  However, in some cases, a Gaussian approximation of the conditional PDF allows to only compute the first two conditional moments, \textit{i.e.} the mean $\widehat{\mathbf{x}}_{k|k} = \mathscr{E}[\mathbf{x}_k|\mathbf{y}^k]$ and the error covariance matrix $\mathbf{P}_{k|k} = cov[\mathbf{x}_k|\mathbf{y}^k]$ which results in $p(\mathbf{x}_k|\mathbf{y}^k) \approx \mathscr{N}\{\mathbf{x}_k;\widehat{\mathbf{x}}_{k|k},\mathbf{P}_{k|k} \}$.

By assuming Gaussian white noise vectors, the prediction step (state prediction) and correction step (measurement update) are carried out via integrating a nonlinear function concerning a normal distribution, or
\textcolor{black}{{\small
\begin{align}
			\widehat{\mathbf{x}}_{k+1|k}\! & =\! \mathscr{E}[\mathbf{x}_{k+1}|\mathbf{y}^k] \!\! =\!\! \int_{\mathbb{R}_n}\!\!\!\!\! \mathbf{f}(\mathbf{x}_k,\mathbf{u}_k) p(\mathbf{x}_{k}|\mathbf{y}^k) \text{d}\mathbf{x}_k\nonumber\\
		    \!\!\! & \approx\!\!\! \int_{\mathbb{R}_n}\!\!\!\!\! \mathbf{f}(\mathbf{x}_k,\mathbf{u}_k) \mathscr{N}\{\mathbf{x}_k;\widehat{\mathbf{x}}_{k|k},\mathbf{P}_{k|k} \}\text{d}\mathbf{x}_k,\label{eq:integrals1}\\
    \widehat{\mathbf{y}}_{k+1|k} \! & = \! \mathscr{E}[\mathbf{y}_{k+1}|\mathbf{x}_{k+1}]\!\! =\!\! \int_{\mathbb{R}_n}\!\!\!\!\!\! \mathbf{h}(\mathbf{x}_{k+1},\mathbf{u}_{k+1}) p(\mathbf{y}_{k+1}|\mathbf{x}_{k+1}) \text{d}\mathbf{x}_{k+1}\nonumber\\
			\!\!\! & \approx\!\!\! \int_{\mathbb{R}_n}\!\!\!\!\! \mathbf{h}(\mathbf{x}_{k+1},\mathbf{u}_{k+1}) \mathscr{N}\{\mathbf{x}_{k+1};\widehat{\mathbf{x}}_{k+1|k},\mathbf{P}_{k+1|k} \}\text{d}\mathbf{x}_{k+1}.
			\label{eq:integrals2}
		\end{align}
	}}
\noindent The third-degree spherical-radial rule is utilized to compute the numerical approximation of the moment integrals (\ref{eq:integrals1}) and (\ref{eq:integrals2}). Next, for an arbitrary function $g(\mathbf{x})$ with $\boldsymbol{\Sigma}$ as the covariance of $\mathbf{x}$, the integral
\begin{small}
		\beq
			\!\!\! I(g) \!= \! \sqrt{2\pi} \vert\boldsymbol{\Sigma} \vert^{-\frac{1}{2}}\int_{\mathbb{R}^n}\!\!\! g(\mathbf{x})exp\left[-\dfrac{1}{2}(\mathbf{x}-\boldsymbol{\mu})^\text{T} \boldsymbol{\Sigma}^{-1}(\mathbf{x}-\boldsymbol{\mu}) \right]\text{d}\mathbf{x},
		\eeq
\end{small}\noindent in the spherical coordinate system becomes 
		\beq
			I(g) = (2\pi)^{-\frac{n}{2}} \int_{r=0}^\infty \int_{\mathbb{U}_n} g(\mathbf{C}r\mathbf{z}+\boldsymbol{\mu})\text{d}\mathbf{z}\, r^{n-1}e^{-\frac{r^2}{2}} \text{d}r,
		\eeq
	where $\mathbf{x} = \mathbf{C}r\mathbf{z}+\boldsymbol{\mu}$ with $\Vert\mathbf{z} \Vert =1$, $\boldsymbol{\mu}$ is the mean and $\mathbf{C}$ is the Cholesky factor of the covariance, $\boldsymbol{\Sigma}$, and $\mathbb{U}_n$ is the unit sphere. Then, we used the symmetric spherical cubature rule to further approximate the integral as
		\beq
			I(g) = \dfrac{1}{2n} \sum\limits_{i=0}^{2n} g(\sqrt{n}(\mathbf{C}{\xi}_i+\boldsymbol{\mu})),
		\eeq
	where ${\xi}_i$ denotes the $i$th cubature point at the intersection of the unit sphere and its axes. The main benefit of this scheme is that the cubature points are obtained off-line using a third-degree cubature rule \cite{liu2014adaptive}. We follow the steps introduced next to compute the estimation of the state vector via the SRCKF algorithm.

\subsubsection*{SRCKF algorithm}
\begin{enumerate}
	\item {\bf Initialization}: The state initial condition is given by $\mathbf{x}_{0|0}\equiv \mathbf{x}_{0}$ with $\widehat{\mathbf{x}}_0=\mathscr{E}[\mathbf{x}_0]$ where the initial covariance matrix is $\mathbf{P}_{0|0}$. We decompose it as $\mathbf{P}_{0|0} = \mathbf{S}_{0|0}\mathbf{S}_{0|0}^\text{T}$ through the Cholesky factorization, \textit{i.e.}
		\benn
			\mathbf{S}_{0|0} = chol\{[\mathbf{x}_0 - \widehat{\mathbf{x}}_0][\mathbf{x}_0 - \widehat{\mathbf{x}}_0]^\text{T}\}.
		\eenn
		Then, generate the cubature points, $\boldsymbol{\xi}_i$, for the initia state vector and the fixed weights, $w_i = w = \dfrac{1}{2n}$, for $i=1,2,\ldots,2n$.

	\item \textbf{Time update (Prediction)} $(k=1,2,\ldots,k_f)$:
		\begin{enumerate}
			\item Evaluation of the cubature points
				\beq
					\mathbf{X}_{i,k-1|k-1} = \mathbf{S}_{k-1|k-1} \boldsymbol{\xi}_i + \widehat{\mathbf{x}}_{k-1|k-1}.
				\eeq
			\item Evaluation of the propagated cubature points via the system dynamics
				\beq
					\mathbf{X}_{i,k|k-1}^* = \mathbf{f}_k(\mathbf{X}_{i,k-1|k-1},\mathbf{u}_{k-1}).
				\eeq
			\item Evaluation of the predicted states based on the generated weights and propagated points
				\beq
					\widehat{\mathbf{x}}_{k|k-1} = \sum\limits_{i=1}^{2n} w_i \mathbf{X}_{i,k|k-1}^*.
				\eeq
			\item Evaluation of the square root of the covariance matrix of the predicted state error covariance
				\beq
					\mathbf{S}_{k|k-1} = triangle\big\{[\boldsymbol{\chi}_{k|k-1}^*, \mathbf{S}_{\mathbf{Q}_{k-1}} ]\big\},
					\label{eq:S}
				\eeq
			where $\boldsymbol{\chi}_{k|k-1}^*$ is a centered weighted matrix, \textit{i.e.}
				\begin{align}
					& \boldsymbol{\chi}_{k|k-1}^* = \dfrac{1}{\sqrt{2n}}[\mathbf{X}_{1,k|k-1}^* - \widehat{\mathbf{x}}_{k|k-1}\nonumber\\
										&\quad \begin{array}{ccc}
															 \mathbf{X}_{2,k|k-1}^* - \widehat{\mathbf{x}}_{k|k-1} &\cdots & \mathbf{X}_{2n,k|k-1}^* - \widehat{\mathbf{x}}_{k|k-1}					
\end{array}],
				\end{align}
				and $\mathbf{S}_{\mathbf{Q}_{k-1}}$ is the square-root of the the process noise such that $\mathbf{Q}_{k-1}=\mathbf{S}_{\mathbf{Q}_{k-1}} \mathbf{S}_{\mathbf{Q}_{k-1}}^\text{T}$.
				Moreover, $\mathbf{B}=triangle\{\mathbf{A}\}$ stands for a general triangularization algorithm, \textit{e.g.} QR decomposition, where $\mathbf{B}$ is a lower triangular matrix. If $\mathbf{C}$ is an upper triangular matrix obtained through the QR decomposition of $\mathbf{A}^\text{T}$, then the lower triangular matrix is given by $\mathbf{B} = \mathbf{C}^{\text{T}}$.
		\end{enumerate}
		
	\vspace{1mm}	
	\item \textbf{Measurement update (Correction)} $(k=1,2,\ldots,k_f)$:
		\begin{enumerate}
			\item Evaluation of the cubature points using the predicted square root matrix, $\mathbf{S}_{k|k-1}$,
				\beq
					\mathbf{X}_{i,k|k-1} = \mathbf{S}_{k|k-1} \boldsymbol{\xi}_ i + \widehat{\mathbf{x}}_{k|k-1}.
				\eeq
			\item Evaluation of the propagated cubature point via the output dynamics
				\beq
					\mathbf{Y}_{i,k|k-1} = \mathbf{h}(\mathbf{X}_{i,k|k-1},\mathbf{u}_k).
				\eeq
			\item Estimation of the predicted measurement vector
				\beq
					\widehat{\mathbf{y}}_{k|k-1} = \sum\limits_{i=1}^{2n} w_i \mathbf{Y}_{i,k|k-1}.
				\eeq
			\item Evaluation of the square root of the innovation covariance matrix
				\beq
					\mathbf{S}_{yy,k|k-1} = triangle\big\{[\mathbf{Y}_{k|k-1}, \mathbf{S}_{\mathbf{R}_{k}}]\big\},			
				\eeq
			where $\mathbf{Y}_{k|k-1}$ is a centered weighted matrix, \textit{i.e.}
				\begin{align}
					& \mathbf{Y}_{k|k-1} = \dfrac{1}{\sqrt{2n}}[\mathbf{Y}_{1,k|k-1} - \widehat{\mathbf{y}}_{k|k-1}\nonumber\\
															& \quad	\begin{array}{ccc}
																\mathbf{Y}_{2,k|k-1} - \widehat{\mathbf{y}}_{k|k-1} &\cdots & \mathbf{Y}_{2n,k|k-1} - \widehat{\mathbf{y}}_{k|k-1}					
\end{array}].
				\end{align}
				$\mathbf{S}_{\mathbf{R}_{k}}$ is also the square-root of the the measurement noise such that $\mathbf{R}_{k}=\mathbf{S}_{\mathbf{R}_{k}} \mathbf{S}_{\mathbf{R}_{k}}^\text{T}$.	
				\vspace{1mm}					
			\item Evaluation of the cross-covariance matrix
				\beq
					\mathbf{P}_{xy,k|k-1} = \boldsymbol{\chi}_{k|k-1}\mathbf{Y}_{k|k-1}^\text{T},
				\eeq
			with the centered weighted matrix $\boldsymbol{\chi}_{k|k-1}$ given by
				\begin{align}
					& \boldsymbol{\chi}_{k|k-1} = \dfrac{1}{\sqrt{2n}}[\mathbf{X}_{1,k|k-1} - \widehat{\mathbf{x}}_{k|k-1} \nonumber\\
					&\quad \begin{array}{ccc}
																\mathbf{X}_{2,k|k-1} - \widehat{\mathbf{x}}_{k|k-1} &\cdots & \mathbf{X}_{2n,k|k-1} - \widehat{\mathbf{x}}_{k|k-1}					
\end{array}].
				\end{align}
			\item Evaluation of the SRCKF filter gain
				\beq
					\mathbf{W}_k = \mathbf{P}_{xy,k|k-1} \mathbf{S}_{yy,k|k-1}^{-\text{T}} \mathbf{S}_{yy,k|k-1}^{-1}. 
				\eeq
			\item Evaluation of the corrected state update based on the measurement
				\beq
					\widehat{\mathbf{x}}_{k|k} = \widehat{\mathbf{x}}_{k|k-1} + \mathbf{W}_k (\mathbf{y}_k - \widehat{\mathbf{y}}_{k|k-1}).
				\eeq
			\item Evaluation of the square-root of the corrected error covariance matrix
				\beq
					\mathbf{S}_{k|k} = triangle\big\{[\boldsymbol{\chi}_{k|k-1}-\mathbf{W}_k\mathbf{Y}_{k|k-1}, \mathbf{W}_k\mathbf{S}_{\mathbf{R}_k}]\big\}.
				\eeq
		\end{enumerate}
\end{enumerate}
\noindent The state estimation process continues iteratively from the second step of the algorithm, \textit{i.e.} the time update (prediction) by setting $k = k + 1$.
\subsection{Multiple-Model SRCKF for Input Delay Estimation}
Time delay estimation introduces a challenge in the parameter identification framework since the variable delay is not transformable to an equivalent random walk process. Rational approximations of the delay such as Pad\'{e} approximation can be considered as alternative solutions; however, the introduced truncation error may be significant and problematic, especially for large and time-varying delays. Thus, to obtain a more accurate delay estimation, the aforementioned SRCKF algorithm is equipped with a multiple-model (MM) framework cascaded with a hypothesis testing module \cite{hanlon2000multiple}.

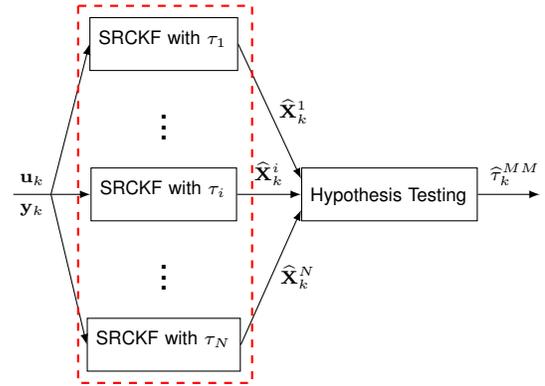
\begin{figure}[!t]
	\centering
		\begin{tikzpicture}[auto, font = {\sf \scriptsize}, cross/.style={path picture={\draw[black] (path picture bounding box.south east) -- (path picture bounding box.north west) (path picture bounding box.south west)-- (path picture bounding box.north east);}}, node distance=2cm,>=latex]


		\node at (-5,0)[input, name=input1] {};

		\node at (-4.5,0)[input, name=inputm] {};		

		\node at (-3,1)[tip, name=input2] {\Large $\mathbf{\vdots}$};

		\node at (-3,-1)[tip, name=input3] {\Large $\mathbf{\vdots}$};		

		\node at (-3,0)[block,text width=1.7cm,text height=.0em,text depth=0em] (CKFi) {\centering SRCKF with $\tau_i$};

 	    \node at (-3,2)[block,text width=1.73cm,text height=.0em,text depth=.0em] (CKF1) {\centering SRCKF with $\tau_1$};

 	    \node at (-3,-2)[block,text width=1.8cm,text height=.0em,text depth=.0em] (CKFN) {\centering SRCKF with $\tau_N$};

		\node at (0,0)[block,text width=2.1cm,text height=.7em,text depth=.2em] (HT) {\centering Hypothesis Testing};
	    \node at (2,0)[output, node distance=2.1cm] (output1) {};
	    \draw [-] (input1) -- node[name=in1tosum1, pos=0.5, above] {$\mathbf{u}_k$} node[name=ms2, pos=.5, below] {$\mathbf{y}_k$} (inputm);

		\draw [->] (inputm) -- node[name=sum1tosum2, pos=0.4] {} (CKF1.west);	 	    

	    \draw [->] (inputm) -- node[name=sum2toCont, pos=0.4] {} (CKFi);

	    \draw [->] (inputm) -- node[name=sum2toCont, pos=0.4] {} (CKFN.west);

	    \draw [->] (CKF1.east) -- node[name=sum2toCont, pos=0.5, right] {$\widehat{\mathbf{X}}_k^1$} (HT.170);

	    \draw [->] (CKFi) -- node[name=sum2toCont, pos=0.5, above] {$\widehat{\mathbf{X}}_k^i$} (HT);

	    \draw [->] (CKFN.east) -- node[name=sum2toCont, pos=0.5, right] {$\widehat{\mathbf{X}}_k^N$} (HT.190);

    		\draw [->] (HT) -- node[name=sum2toCont, pos=0.6, above] {$\widehat{\tau}_k^{MM}$} (output1);

    		\draw[red,thick,dashed] ($(CKF1.north west)+(-0.15,.15)$)  rectangle ($(CKFN.south east)+(0.15,-0.15)$);

	\end{tikzpicture}

    \caption{Bank of $N$ parallel SRCKFs for delay estimation}

    \label{fig:Bank}

\end{figure}

The underlying idea of the MMSRCKF method is to use a bank of $N$ identical SRCKFs in a parallel setting, as shown in Fig. \ref{fig:Bank}. Every SRCKF uses the same measurement and input data, but a different delay is assigned to each filter. The $i$th element in the bank provides us with a state vector estimation $\mathbf{X}_k^i$ together with the residuals $\mathbf{r}_k^i = \mathbf{y}_k - \widehat{\mathbf{y}}_k^i$. By having this information, a hypothesis testing block can then be used to estimate the value of the delay. Specifically, if the delay matches the one assigned to the $i$th SRCKF element, then the corresponding residual is essentially a zero-mean white noise process, \textit{i.e.} $\mathscr{E}[\mathbf{r}_k^i] = 0$, and its covariance is given by
	\beq
		\mathscr{E}[\mathbf{r}_k^i (\mathbf{r}_k^i)^\text{T}] = \mathbf{H}\mathbf{P}_k^i \mathbf{H}^\text{T} + \mathbf{R} \triangleq \mathbf{R}_k^i,
	\eeq
	where $\mathbf{H}=[1 \; 0 \; 0 \; 1]$, $\mathbf{P}_k^i$ denotes the estimation covariance at the $k$th step, and $\mathbf{R}$ denotes the measurement noise covariance.
The conditional probability density function of the $i$th SRCKF element measurement can be computed through
	\beq
		f(\widehat{y}_k^i|y_k) = \dfrac{1}{(2\pi)^{\frac{m}{2}}\vert \mathbf{R}_k^i\vert^{\frac{1}{2}}}exp\Big\{-\dfrac{1}{2}(\mathbf{r}_k^i)^\text{T} (\mathbf{R}_k^i)^{-1} \mathbf{r}_k^i \Big\},
	\eeq
where $m$ is the dimension of available measurements at each time step. Then, the conditional probability of each hypothesis is
	\beq
		p_k^i = \dfrac{f(\widehat{y}_k^i|y_k)p_{k-1}^i}{\sum\limits_{j=1}^N f(\widehat{y}_k^j|y_k)p_{k-1}^j},
	\eeq
where $p_k^i$ can be interpreted as the normalized conditional probability of the case when the delay equals the assigned value to the $i$th filter, \textit{i.e.} $\sum\limits_{j=1}^N p_{k}^j=1$. Now, it is possible to estimate the delay according to the filter, which has the highest probability. However, to obtain a more accurate delay estimation and to avoid large fluctuations, instead of choosing the block with most likely delay estimation, we treat the hypotheses resulting as weights and blend them to improve the delay estimation. In other words, we can estimate the time delay as
	\beq
		\hat{\tau}_k^{MM} = \sum\limits_{j=1}^N p_{k}^j \tau_k^j,
	\eeq	
\noindent where $\tau_k^j$ is the delay estimation of the $i$th filter. In the next section, we will present the mathematical model describing the dynamics of the MAP response to the PHP drug infusion.

\section{MAP Response Modeling} \label{sec:ProbForm}
\begin{figure}[!t] 
\hspace*{-.1in}
\centering \includegraphics[width=1.05\columnwidth, height=2.00in]{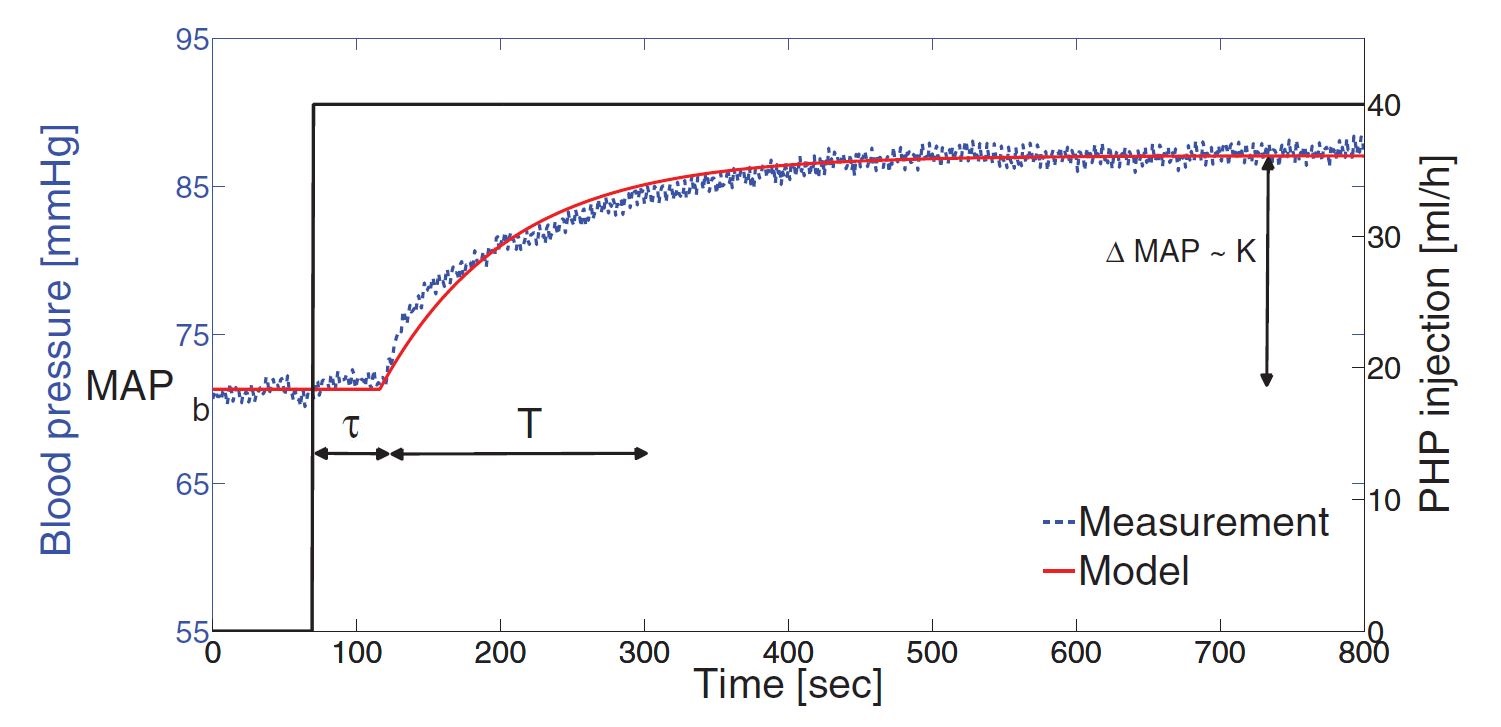} 
\caption{Typical MAP variations in response to PHP step injection \cite{luspay2016adaptive}} 
\label{fig:MAPresponse}
\end{figure}

 The following first-order model with an input delay has been broadly considered and implemented in the literature to characterize the patient's MAP response to the infusion of a vasoactive drug, such as phenylephrine (PHP) \cite{sandu2016reinforcement, luspay2016adaptive, StasoujianRobust}:
\beq
T(t) \cdot \dot{\Delta MAP}(t) + \Delta MAP(t) = K(t) \cdot u(t-\tau(t)),
\label{eq:MAP response TF}
\eeq
where  $\Delta MAP(t)$  stands for the MAP changes in $mmHg$ from its baseline value, \textit{i.e.} $\Delta MAP (t)= MAP (t) -MAP_b(t)$, $u(t)$ is the drug injection rate in $ml/h$, $K(t)$ denotes the patient's sensitivity to the administered drug, $T(t)$ is the lag time describing the uptake, distribution and biotransformation of the drug \cite{isaka1993control}, and $\tau(t)$ is the time delay for the drug to reach the circulatory system from the injection site. This model structure seems to adequately describe a patient's physiological response to the PHP drug injection. Fig. \ref{fig:MAPresponse} presents a typical MAP response due to a step PHP infusion versus a matched response of (\ref{eq:MAP response TF}). This figure also illustrates the interpretation of the model parameters $K(t)$, $T(t)$, $\tau(t)$, $MAP_b(t)$ which have been obtained to fit the MAP response using a least-squares optimization method.  Although the proposed model structure (\ref{eq:MAP response TF}) is qualitatively able to represent the characteristics of the MAP response to the infusion of PHP, experiments show that the model parameters vary significantly over time due to patients' pharmacological variability subject to the vasoactive drug infusion. That is, the model parameters and delay could vary remarkably from patient-to-patient (inter-patient variability), as well as, for a given patient over time (intra-patient variability) \cite{isaka1993control, rao2003experimental}. 
For the implementation of recursive sequential estimation tools, we discretize the continuous-time model (\ref{eq:MAP response TF}) at the sampling rate of $T_s$ as follows
\beq
	\left\{\begin{array}{l}
		x_{k+1} = \big(1-\dfrac{T_s}{T_k} \big) x_k + \dfrac{K_kT_s}{T_k} u_{(k-\frac{\tau_k}{T_s})},\\ [5pt]
		y_k = x_k + MAP_{b_k},
		\end{array}\right.
\label{eq:DiscretizedEq}
\eeq 
\noindent where $x_k = \Delta MAP_k = MAP_k - MAP_{b_k}$ at the $k$th time instant.  In (\ref{eq:DiscretizedEq}), we augment the state vector with the parameters to be estimated, namely $K_k, T_k,$ and $MAP_{b_k}$ by assuming local random-walk dynamics. In other words
\beq
	\mathbf{X}_k^\text{T}\!=\! [{X}_k^1 \,\, {X}_k^2 \,\, {X}_k^3 \,\, {X}_k^4]\! =\! [{\Delta MAP}_k \,\, K_k \,\, T_k \,\, MAP_{b_k}].
\eeq
Since model parameters are all time-varying and assumed to be \textit{a priori} unknown, (\ref{eq:DiscretizedEq}) represents a nonlinear equation with regards to the state vector, $\mathbf{X}_k$, that can be expressed as the following nonlinear dynamics
\beq
	\left\{\begin{array}{l}
	X_{k+1}^1 = \mathbf{f}_k(\mathbf{X}_k,u_k) + w_k,\\[3pt]
	y_k = h_k(\mathbf{X}_k) + v_k,
	\end{array}\right.
\eeq
with 
\beq
	\left\{\begin{array}{l}
	f_k^1(\mathbf{X}_k,u_k) = \big(1-\dfrac{T_s}{X_k^3} \big) X_k^1 + \dfrac{T_s X_k^2}{X_k^3} u_{(k-\frac{\tau_k}{T_s})},\\[6pt]
	h_k(\mathbf{X}_k) = X_k^1 + X_k^4.
\end{array}			 \right.
\eeq
The process noise, $w_k$, and the measurement noise, $v_k$, are both assumed to be additive and statistically independent zero-mean Gaussian processes with covariances given by $\mathbf{Q}_{k}$ and ${R}_{k}$, respectively. Although such an augmentation facilitates the estimation procedure, the time-varying input delay neither can be included in the augmented state vector nor be captured by a random walk process. Thus, the time delay is estimated through a multiple-model (MM) hypothesis testing process along with the SRCKF, discussed in Section \ref{sec:methodology}. 

Next, we test the proposed MMSRCKF estimation algorithm in a simulation where the patient's model parameters are generated by nonlinear functions based on clinical observations. Then, we validate the verified estimation framework using the experimental data from animal experiments. 

\section{MAP Response Estimation Results and Validations}
\label{sec:EstimationResults}

\begin{figure}[!t] 
\hspace*{-.17in}
\centering \includegraphics[width=\columnwidth, height=2.3in]{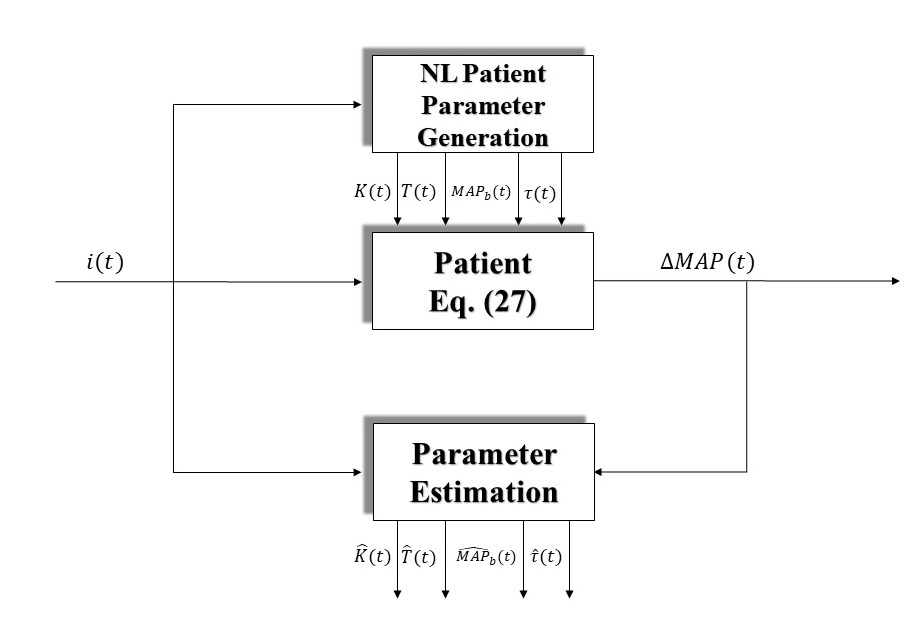} 
\caption{Structure of nonlinear patient parameter generation} 
\label{fig:nlpatientstructure}
\end{figure}

\begin{figure}[!t] 
\hspace*{-.17in}
\centering \includegraphics[width=\columnwidth, height=1.95in]{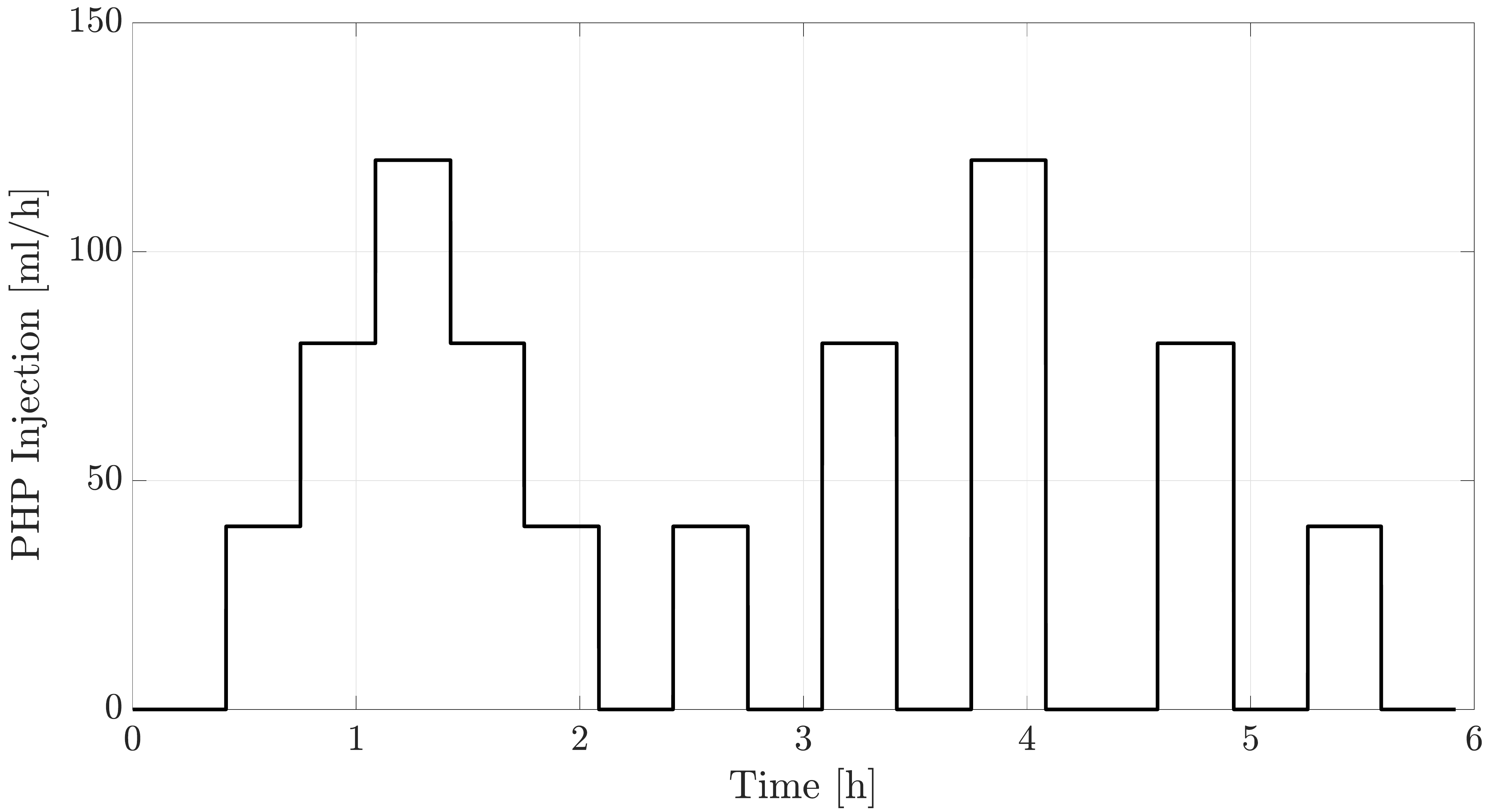}
\caption{Profile of piecewise constant PHP drug injection} 
\label{fig:infusionstep}
\end{figure}

\begin{figure}[!t] 
\hspace*{-.17in}
\centering \includegraphics[width=\columnwidth, height=1.95in]{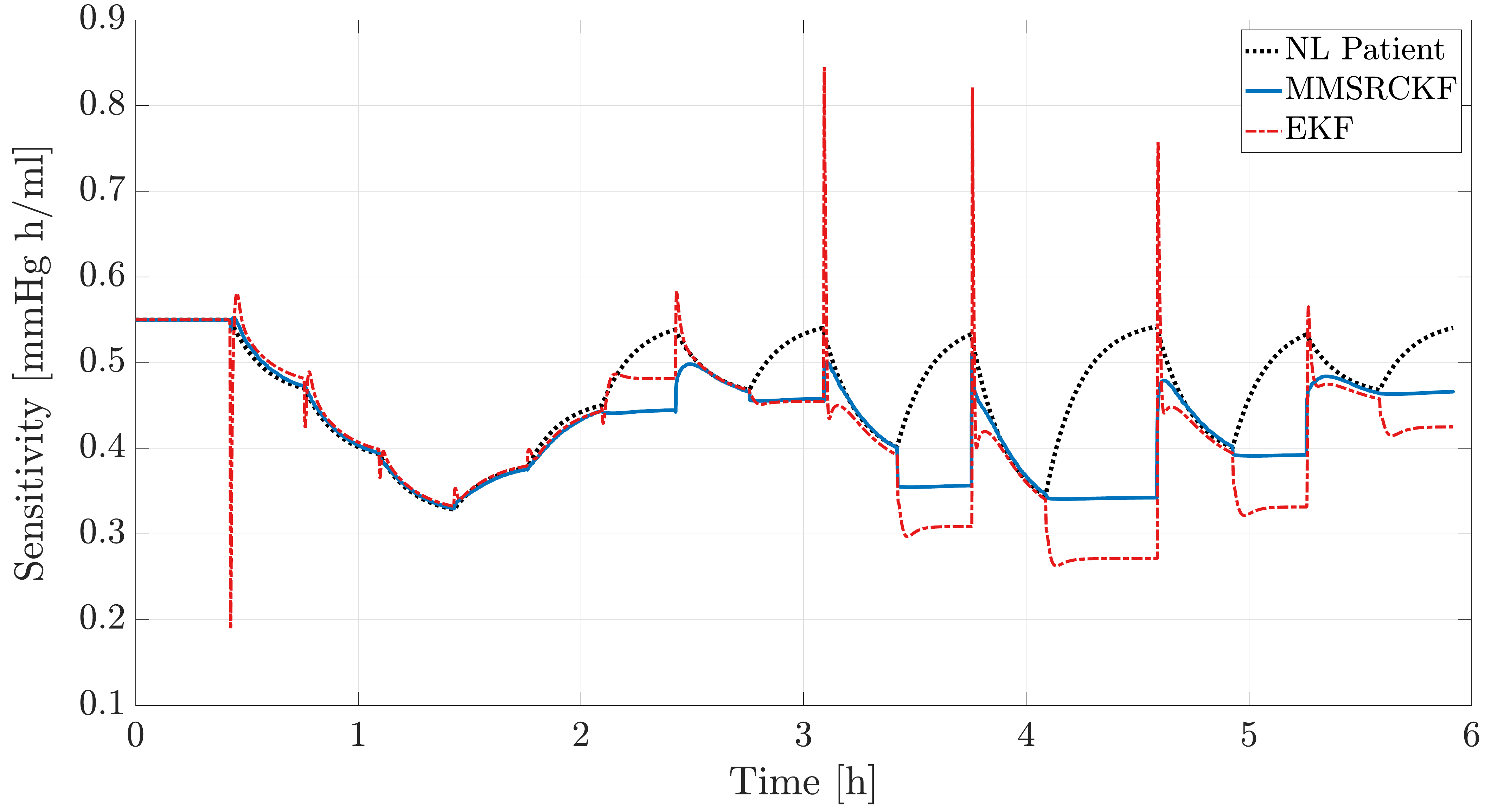}
\caption{Nonlinear patient sensitivity estimation} 
\label{fig:K_est}
\end{figure}

\begin{figure}[!t] 
\hspace*{-.17in}
\centering \includegraphics[width=\columnwidth, height=1.95in]{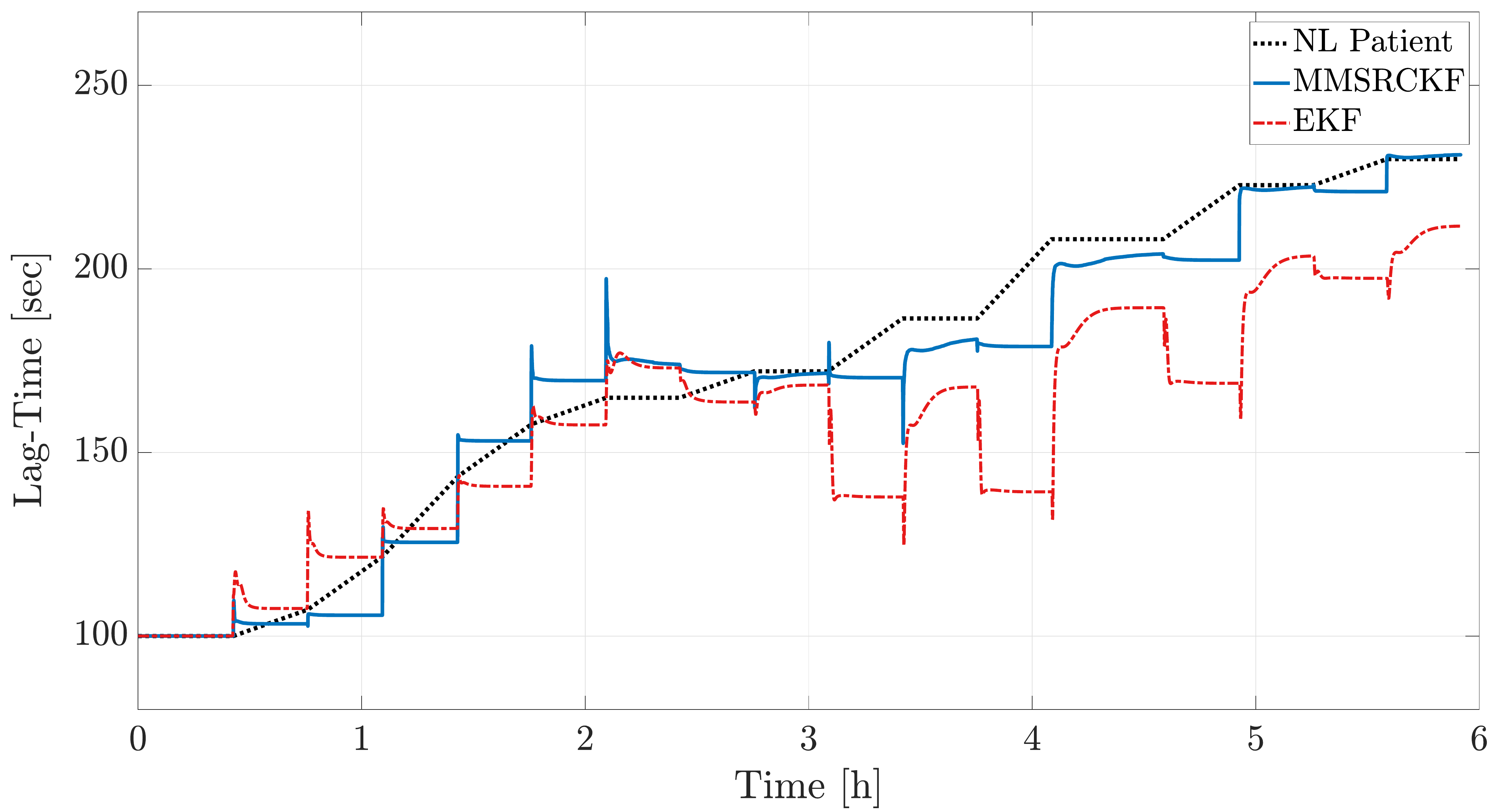} 
\caption{Nonlinear patient lag time estimation} 
\label{fig:lag_est}
\end{figure}

\begin{figure}[!t] 
\hspace*{-.17in}
\centering \includegraphics[width=\columnwidth, height=1.95in]{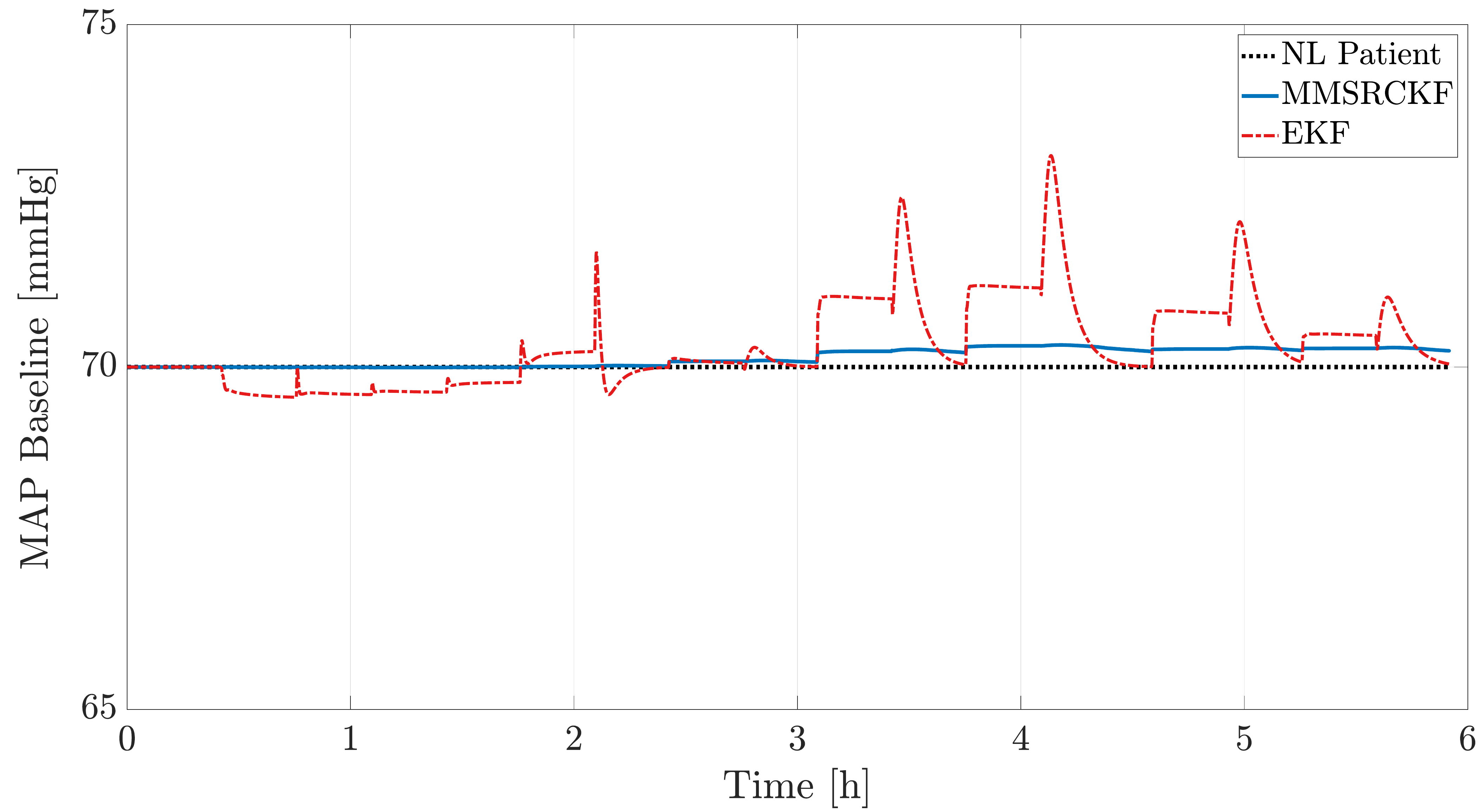} 
\caption{Nonlinear patient baseline MAP estimation} 
\label{fig:baseline_est}
\end{figure}

\begin{figure}[!t] 
\hspace*{-.17in}
\centering \includegraphics[width=\columnwidth, height=1.95in]{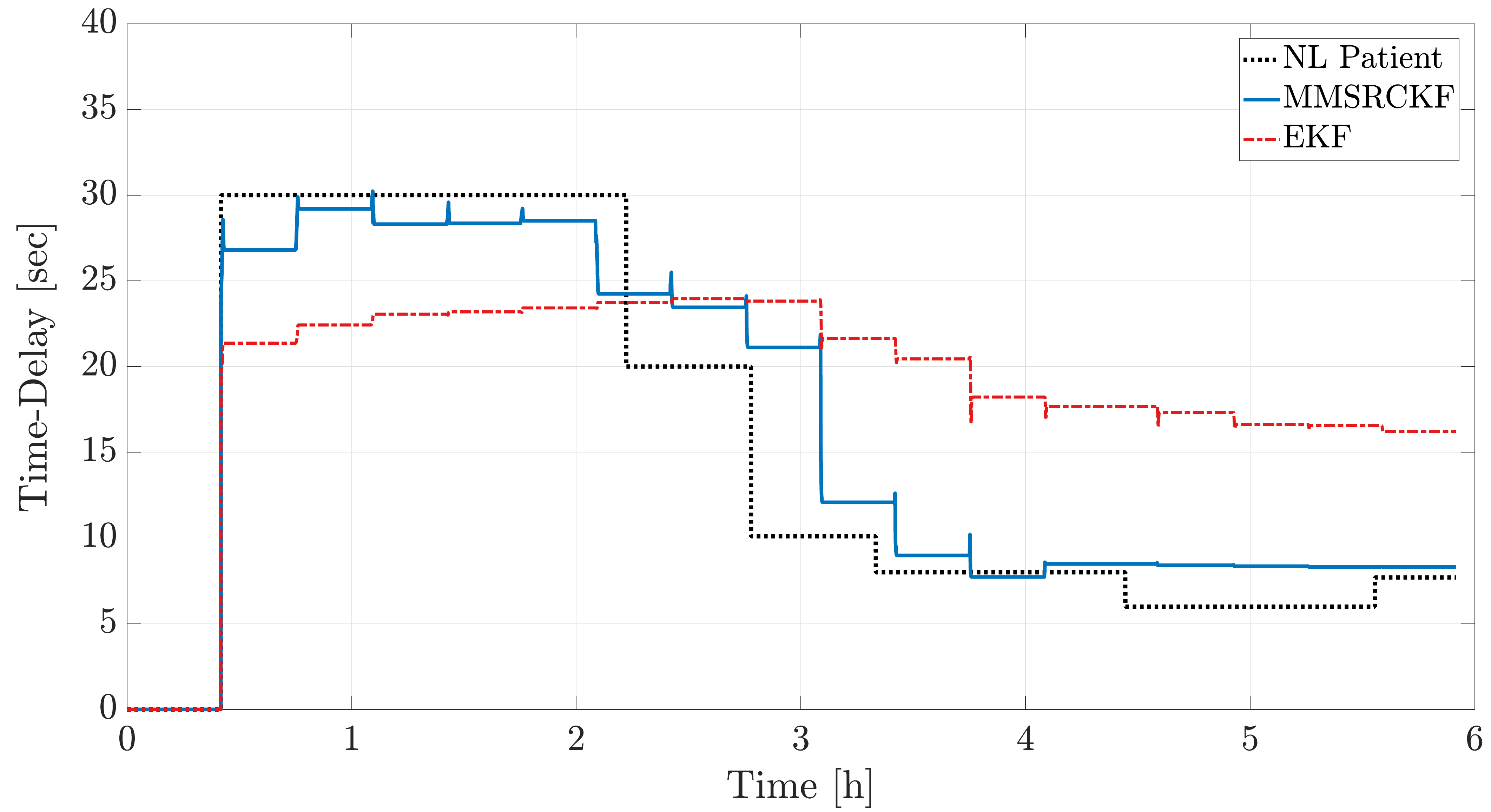} 
\caption{Nonlinear patient input delay estimation} 
\label{fig:delay_est}
\end{figure}

In order to validate the proposed parameter estimation method, first, we need to build a realistic simulation model of an individual's MAP response to the drug infusion with some known model parameters. Adopting (\ref{eq:MAP response TF}), we generate the patient's nonlinear time-varying model parameters, \textit{i.e.} $K(t)$, $T(t)$, $\tau(t)$, and $MAP_b(t)$ based on clinical observations as follows \cite{StasoujianRobust}
\begin{align}
a_k \dot{K}(t) + K(t)  = k_0 exp\{ - k_1 i(t) \},\:\:\:\:\:\:\:\:\:\:\:\:\:\:\:\:\:\:\:\:\:\:\:\:\: \nonumber\\
T(t)= sat_{\:[T_{\min}, T_{\max}]} \: \{b_{T} \int_{0}^{t} i(t) \:dt \},\:\:\:\:\:\:\:\:\:\:\:\:\:\:\:\:\:\:\:\:\:\:\nonumber\\
\begin{cases}a_{\tau,2} \dddot{\tau}(t) + a_{\tau,1} \ddot{\tau}(t) + \dot{\tau}(t) = b_{\tau,1} \dot{i}(t) + i(t), & \:\:\:\:\:\:\: t\geq t_{i_0}, \\\tau(t)=0, & otherwise,\end{cases}
\label{eq:nlpatientparameters}
\end{align}
\noindent where $i(t)$ is the drug injection and $a_k$, $k_0$, $k_1$, $b_{T}$, $a_{\tau,2}$, $a_{\tau,1}$, and $b_{\tau,1}$ are uniformly distributed random coefficients given in Table \ref{tab:coef} \cite{Craig2004}. Also, the MAP baseline value, $MAP_b(t)$, is assumed to be constant and equal to  $70\, mmHg$ for the considered nonlinear patient.
\begin{table}[!t]
\centering
\caption{Probabilistic distributions of coefficients in (\ref{eq:nlpatientparameters}) }
\label{tab:coef}
\begin{tabular}{cc}
\hline
Parameter    & Distribution                             \\ \hline
$a_k$        & $\mathcal{U}(500, 600)$                  \\
$k_0$        & $\mathcal{U}(0.1, 1)$                    \\
$k_1$        & $\mathcal{U}(0.002, 0.007)$              \\
$b_{T}$      & $\mathcal{U}(10^{-4}, 3 \times 10^{-4})$ \\
$a_{\tau,1}$ & $\mathcal{U}(5, 15)$                     \\
$a_{\tau,2}$ & $\mathcal{U}(5, 15)$                     \\
$b_{\tau,1}$ & $\mathcal{U}(80, 120)$                   \\ \hline
\end{tabular}
\end{table}
As per (\ref{eq:nlpatientparameters}), the model parameters $K(t)$, $T(t)$, and $\tau(t)$ are nonlinear functions of the drug infusion rate, $i(t)$. Fig. \ref{fig:nlpatientstructure} demonstrates the general structure of the nonlinear patient parameter generation process. The model parameters are generated based on the given infusion rate, $i(t)$, while the parameter estimation tool estimates the model parameters sub-optimally, using the input drug infusion rate and measured output MAP. Fig. \ref{fig:infusionstep} shows the piecewise constant PHP drug infusion profile, that is used to generate the nonlinear patient parameters.
Using the generated model parameters, we evaluate the performance of the proposed  MMSRCKF method in estimating them, and the estimation results are compared to the previously reported EKF algorithm \cite{luspay2016adaptive}. Figs. \ref{fig:K_est}, \ref{fig:lag_est}, \ref{fig:baseline_est}, and \ref{fig:delay_est} show the estimation results for the model parameters, namely the  sensitivity $K(t)$, time constant $T(t)$, MAP baseline value $MAP_b(t)$, and time delay $\tau(t)$, respectively. As we can see, the implemented MMSRCKF method outperforms the EKF in terms of accuracy and the convergence speed. The MMSRCKF online estimation results show better matches with the generated nonlinear patient reference parameters. It should be noted that the computation complexity of both CKF and EKF algorithms equally grows as $n^3$ where $n$ denotes the system size where the former filter is more accurate and numerically more stable. Table \ref{tab:rmse} further compares the root mean square errors (RMSEs) of the model parameters and estimated MAP response in both algorithms by which the error reduction is obvious using MMSRCKF.
\begin{table}[!t]
\centering
\caption{Estimation root mean square errors (RMSEs)}
\label{tab:rmse}
\begin{tabular}{ccc}
\hline
 & \multicolumn{2}{c}{RMSE} \\ \cline{2-3} 
Parameter & MMSRCKF & EKF \\ \hline
$K$ & 0.061 & 0.095 \\
$T$ & 8.370 & 24.917 \\
$MAP_b$ & 0.188 & 0.706 \\
$\tau$ & 3.128 & 9.114 \\
$MAP$ & 0.008 & 0.202 \\ \hline
\end{tabular}
\end{table}

\begin{figure}[t] 
\hspace*{-.17in}
\centering \includegraphics[width=\columnwidth, height=2.2in]{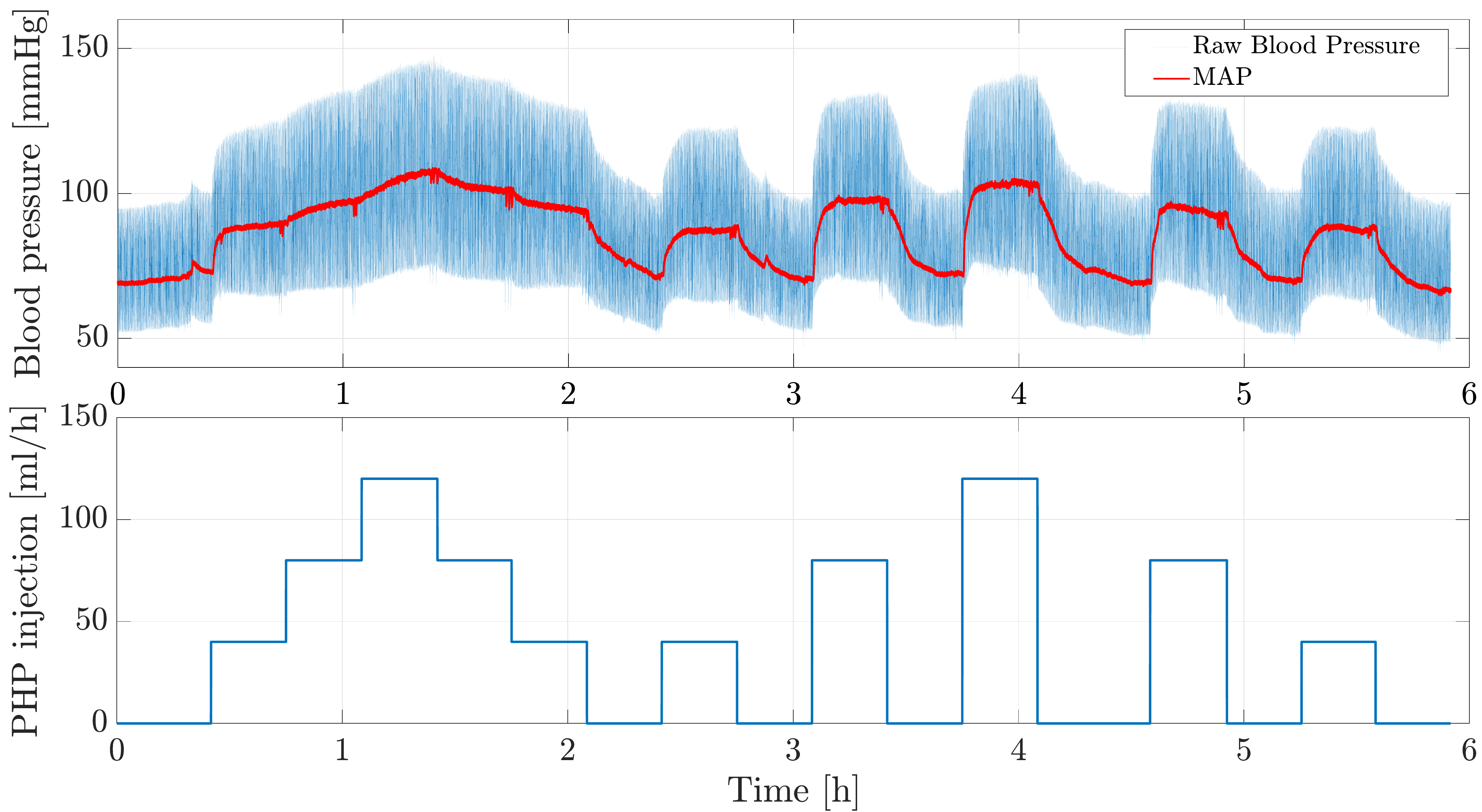} 
\caption{Instantaneous blood pressure and MAP response to a piecewise constant PHP drug injection in animal experiments} 
\label{fig:Bloodpressure}
\end{figure}

\begin{figure}[t] 
\hspace*{-.17in}
\centering \includegraphics[width=\columnwidth, height=1.9in]{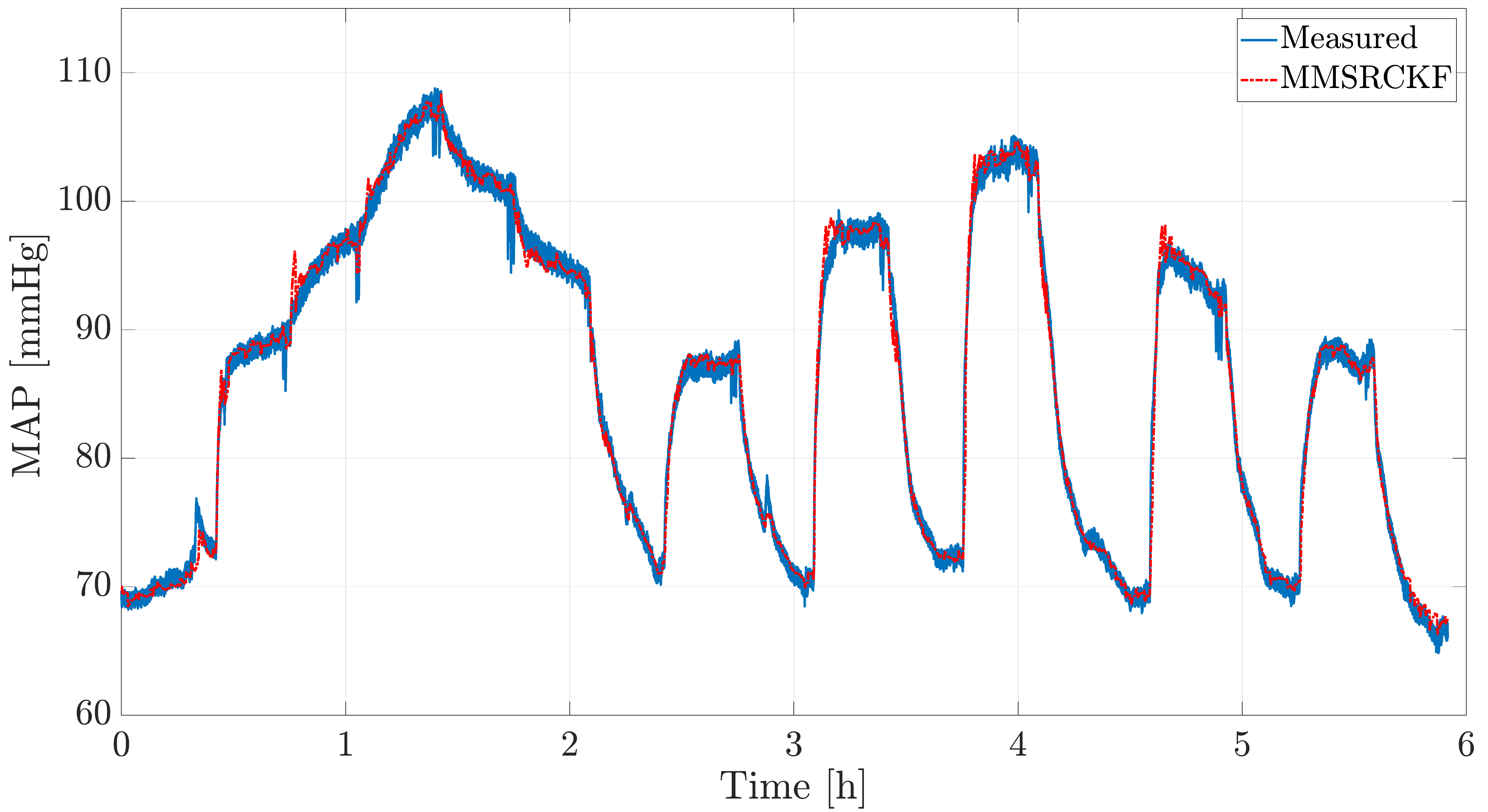} 
\caption{MAP estimation results in animal experiments} 
\label{fig:MAPestimation}
\end{figure}

\begin{figure}[!t] 
\hspace*{-.17in}
\centering \includegraphics[width=\columnwidth, height=1.9in]{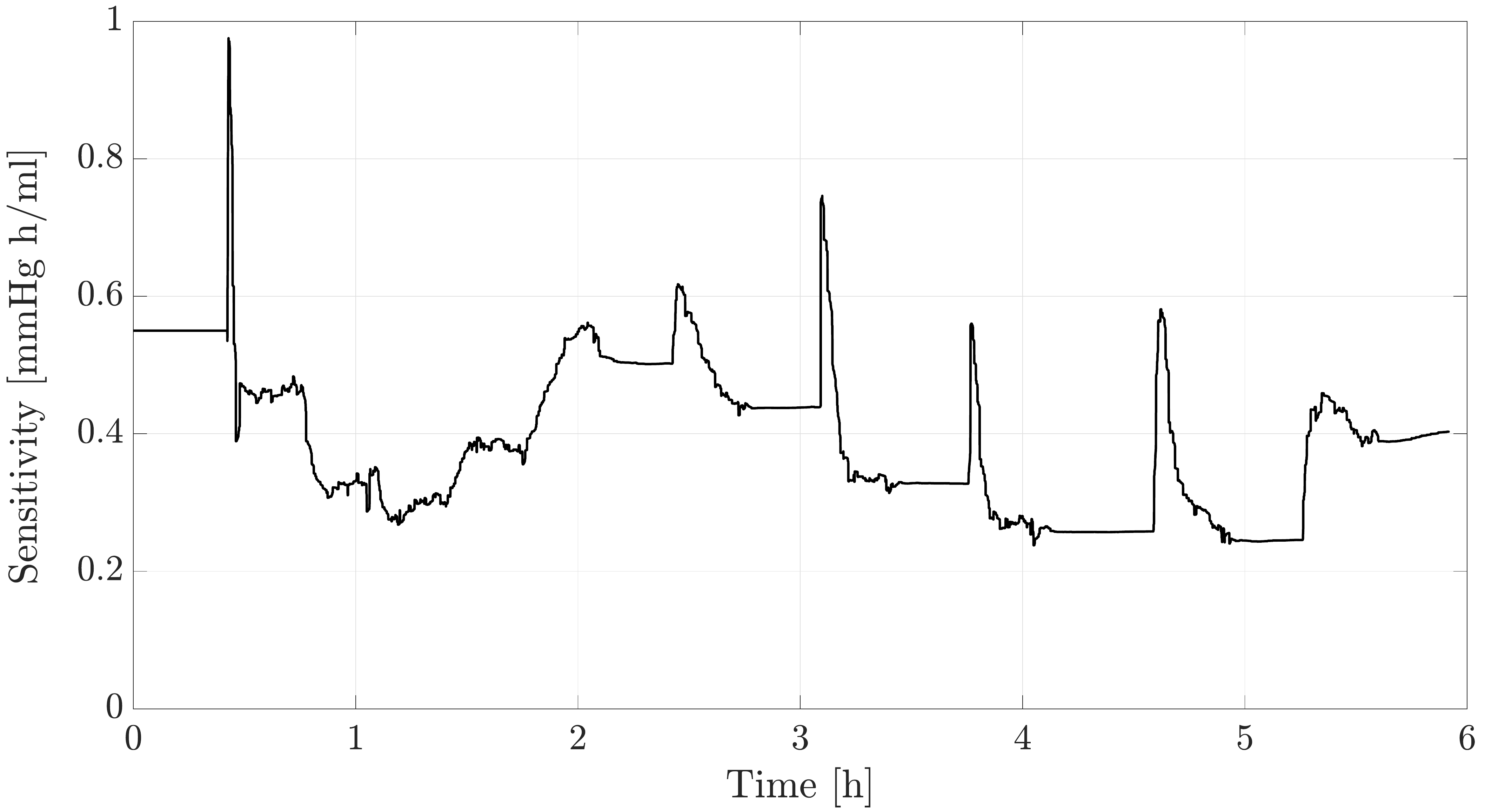} 
\caption{Sensitivity estimation in animal experiments} 
\label{fig:Par_Sensitivity}
\end{figure}

\begin{figure}[!t] 
\hspace*{-.17in}
\centering \includegraphics[width=\columnwidth, height=1.9in]{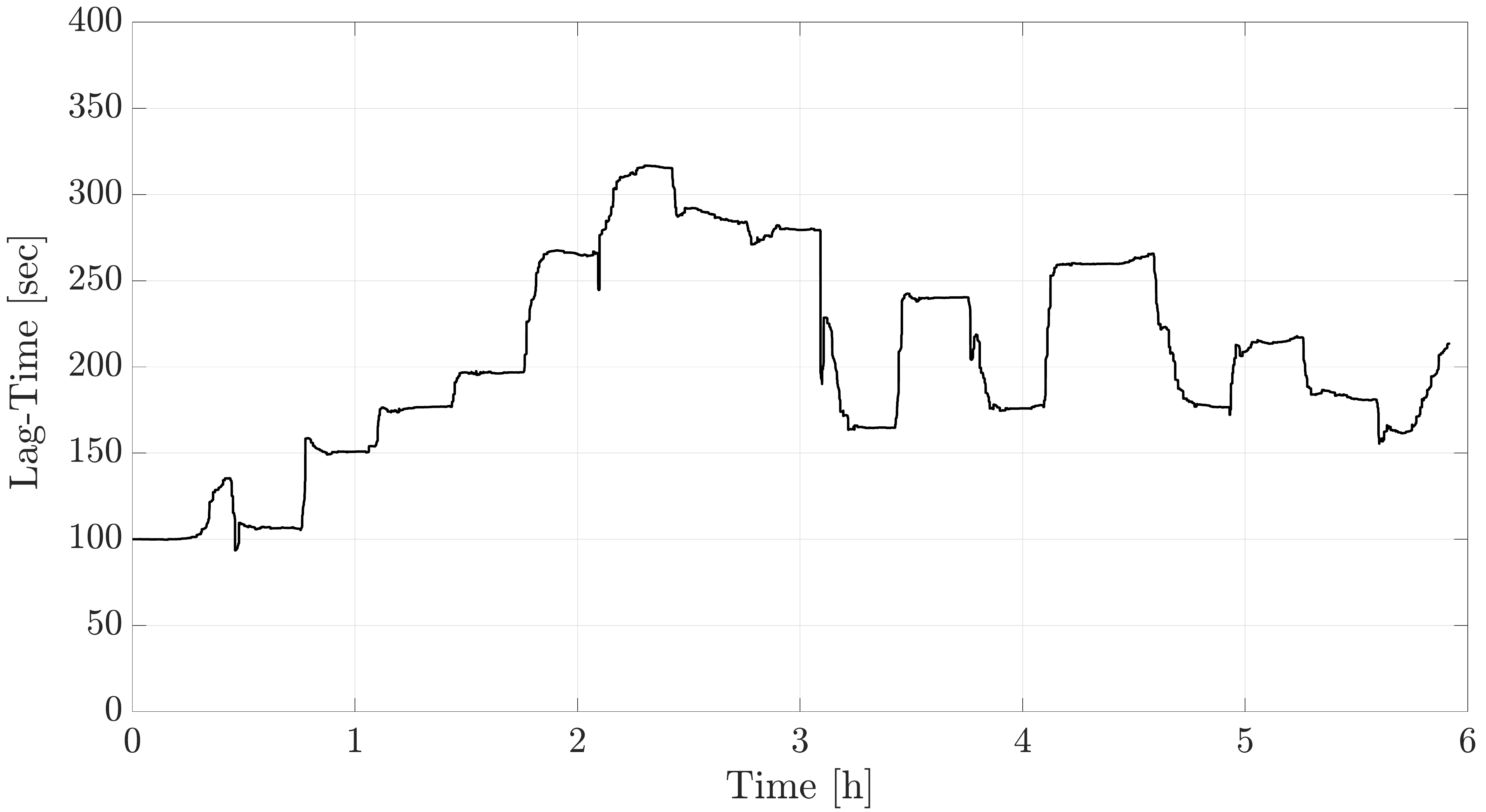} 
\caption{Lag time estimation in animal experiments} 
\label{fig:Par_Lag}
\end{figure}

\begin{figure}[!t] 
\hspace*{-.165in}
\centering \includegraphics[width=0.985\columnwidth, height=1.9in]{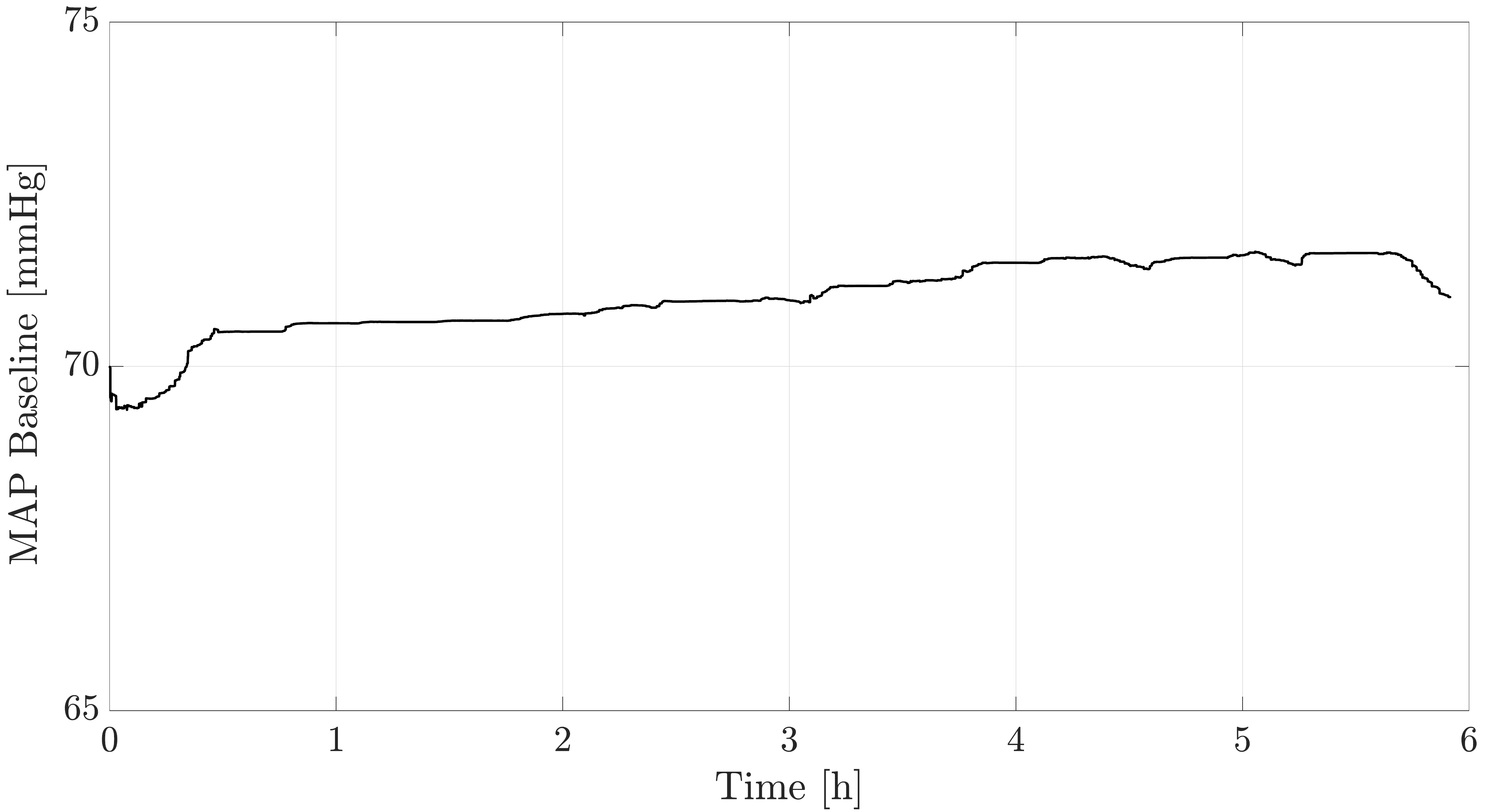} 
\caption{Baseline MAP estimation in animal experiments} 
\label{fig:Par_Baseline}
\end{figure}

\begin{figure}[!t] 
\hspace*{-.165in}
\centering \includegraphics[width=0.985\columnwidth, height=1.9in]{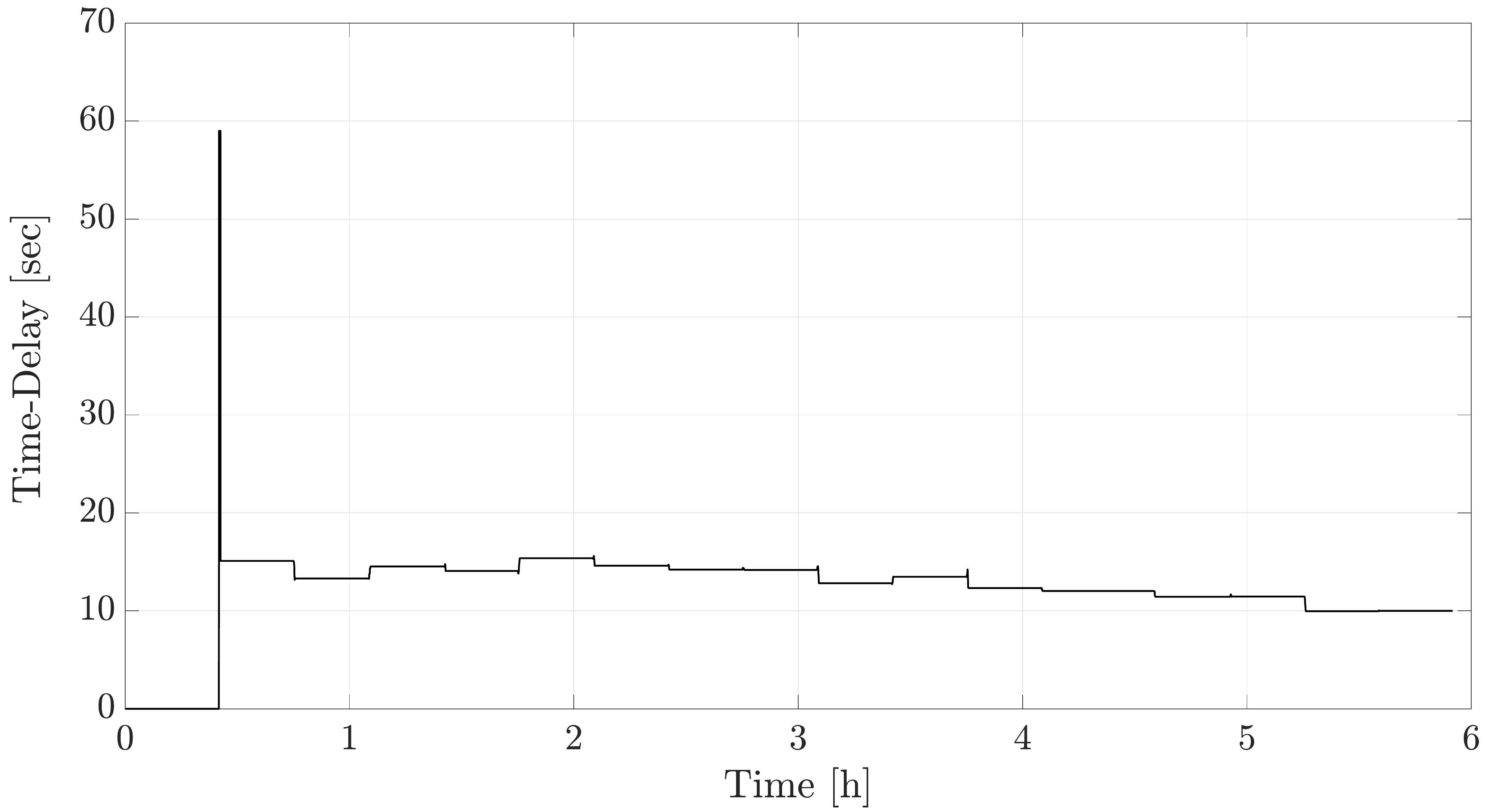} 
\caption{Time delay estimation in animal experiments} 
\label{fig:Par_Delay}
\end{figure}

In the next step, we implemented the MMSRCKF algorithm on the collected data from an actual animal experiment. The input PHP drug infusion rates and output MAP measurements for a $55$ kg anesthetized swine were recorded at the Resuscitation Research Laboratory at the Department of Anesthesiology, UTMB at Galveston, Texas. Precisely speaking, an intramuscular injection of ketamine was used to sedate the swine which were maintained under anesthetic conditions by the continuous infusion of propofol. A Philips MP2 transport device with a sampling frequency of $20\, Hz$ was used to monitor the blood pressure response over a $6$-hour experiment, while the PHP drug was being infused through a bodyguard infusion pump. Fig. \ref{fig:Bloodpressure} shows the piecewise constant PHP drug infusion profile versus the corresponding measured raw blood pressure response and the MAP response over time. We then utilized this dataset for the validation of the estimation of the MAP dynamic model parameters using the proposed MMSRCKF methodology. The experimental dataset was re-sampled at the sampling frequency of $0.2\, Hz$.  
Regarding the multiple-model part of the MMSRCKF algorithm for the delay estimation, the estimation accuracy versus the algorithm speed of convergence triggered a trade-off which needed to be addressed with care; hence, it was essential to choose an appropriate number of the bank of SRCKFs constructing the MMSRCKF structure. In this work, we examined a bank of $11$ SRCKFs with the delay interval of $\tau \in [0\;100] s$. Consequently, the time gridding for the evenly distributed filters was equal to $10 s$. The MAP estimation of the proposed MMSRCKF algorithm, as well as the clinically acquired MAP measurements, are illustrated in Fig. \ref{fig:MAPestimation}, which suggests that the proposed identification method is capable of accurately capturing the MAP response of the swine to the injection of the PHP drug. Additionally, the estimation of the model parameters, namely the sensitivity $K(t)$, time constant $T(t)$, MAP baseline value $MAP_b(t)$, and time delay $\tau(t)$, are depicted in Figs. \ref{fig:Par_Sensitivity}, \ref{fig:Par_Lag}, \ref{fig:Par_Baseline}, and \ref{fig:Par_Delay}, respectively. The estimated parameter values followed the expected trends, as discussed in detail in \cite{StasoujianRobust}. Furthermore, the delay estimation in Fig. \ref{fig:Par_Delay} demonstrated a sharp initialization peak right after the initial injection of the drug and followed a slowly decaying trend during the rest of the experiment as anticipated \cite{Craig2004}. 
\section{Conclusion}\label{sec:conclusion}
Precise estimation of hemodynamics characteristics and mean arterial blood pressure in response to vasoactive drug administration is pivotal to design an effective controller to meet closed-loop physiological response requirements in various clinical scenarios. Real-time estimation of such dynamic models was examined in this paper. Due to the inter- and intra-patient variability, a parameter-varying model with varying input delay was deemed to account for model parameter variations. A Bayesian estimation scheme known as cubature Kalman filter was used because of its convergence speed, nonlinear system handling, and numerical stability. The varying parameters of the nonlinear system corrupted by noise were estimated through the proposed framework. Since the input delay cannot be captured via a random-walk process, the filter was augmented with a multiple-model module. Delay and parameters estimation results in comparison to classical extended Kalman filter were demonstrated which verified the advantage of the utilized Bayesian approach.
\bibliographystyle{IEEEtran}
\bibliography{Ref.bib}

\begin{thebibliography}{10}
\providecommand{\url}[1]{#1}
\csname url@samestyle\endcsname
\providecommand{\newblock}{\relax}
\providecommand{\bibinfo}[2]{#2}
\providecommand{\BIBentrySTDinterwordspacing}{\spaceskip=0pt\relax}
\providecommand{\BIBentryALTinterwordstretchfactor}{4}
\providecommand{\BIBentryALTinterwordspacing}{\spaceskip=\fontdimen2\font plus
\BIBentryALTinterwordstretchfactor\fontdimen3\font minus
  \fontdimen4\font\relax}
\providecommand{\BIBforeignlanguage}[2]{{%
\expandafter\ifx\csname l@#1\endcsname\relax
\typeout{** WARNING: IEEEtran.bst: No hyphenation pattern has been}%
\typeout{** loaded for the language `#1'. Using the pattern for}%
\typeout{** the default language instead.}%
\else
\language=\csname l@#1\endcsname
\fi
#2}}
\providecommand{\BIBdecl}{\relax}
\BIBdecl

\bibitem{Craig2004}
C.~R. Craig and R.~E. Stitzel, \emph{Modern pharmacology with clinical
  applications}.\hskip 1em plus 0.5em minus 0.4em\relax Baltimore, MD, USA:
  Lippincott Williams \& Wilkins, 2004.

\bibitem{DaSilva2019}
S.~J. da~Silva, T.~A. Scardovelli, S.~R.~M. da~Silva~Boschi, S.~C.~M.
  Rodrigues, and A.~P. da~Silva, ``{Simple adaptive PI controller development
  and evaluation for mean arterial pressure regulation},'' \emph{Research on
  Biomedical Engineering}, vol.~35, no.~2, pp. 157--165, 2019.

\bibitem{herget2008approach}
S.~Herget-Rosenthal, F.~Saner, and L.~S. Chawla, ``Approach to hemodynamic
  shock and vasopressors,'' \emph{Clinical Journal of the American Society of
  Nephrology}, vol.~3, no.~2, pp. 546--553, 2008.

\bibitem{StasoujianRobust}
S.~Tasoujian, S.~Salavati, M.~Franchek, and K.~Grigoriadis, ``Robust imc-pid
  and parameter-varying control strategies for automated blood pressure
  regulation,'' \emph{International Journal of Control, Automation and
  Systems}, vol.~17, no.~7, pp. 1803--1813, 2019.

\bibitem{kashihara2004adaptive}
K.~Kashihara, T.~Kawada, K.~Uemura, M.~Sugimachi, and K.~Sunagawa, ``Adaptive
  predictive control of arterial blood pressure based on a neural network
  during acute hypotension,'' \emph{Annals of Biomedical Engineering}, vol.~32,
  no.~10, pp. 1365--1383, 2004.

\bibitem{Arasaratnam2009}
I.~Arasaratnam and S.~Haykin, ``{Cubature Kalman filters},'' \emph{IEEE
  Transactions on Automatic Control}, vol.~54, no.~6, pp. 1254--1269, 2009.

\bibitem{Zhao2018}
X.~Zhao, J.~Li, X.~Yan, and S.~Ji, ``{Robust adaptive cubature Kalman filter
  and its application to ultra-tightly coupled SINS/GPS navigation system},''
  \emph{Sensors}, vol.~18, no. 7:2352, pp. 1--19, 2018.

\bibitem{Furutani2004}
E.~Furutani, M.~Araki, S.~Kan, T.~Aung, H.~Onodera, M.~Imamura, G.~Shirakami,
  and S.~Maetani, ``An automatic control system of the blood pressure of
  patients under surgical operation,'' \emph{International Journal of Control,
  Automation, and Systems}, vol.~2, no.~1, pp. 39--54, 2004.

\bibitem{Gao2005}
Y.~Gao and M.~J. Er, ``An intelligent adaptive control scheme for postsurgical
  blood pressure regulation,'' \emph{IEEE Transactions on Neural Networks},
  vol.~16, no.~2, pp. 475--483, 2005.

\bibitem{Slate1979}
J.~B. Slate, L.~C. Sheppard, V.~C. Rideout, and E.~H. Blackstone, ``A model for
  design of a blood pressure controller for hypertensive patients,'' \emph{IFAC
  Proceedings}, vol.~12, no.~8, pp. 867--874, 1979.

\bibitem{Zhu2008}
K.~Y. Zhu, H.~Zheng, and D.~G. Zhang, ``A computerized drug delivery control
  system for regulation of blood pressure,'' \emph{International Journal of
  Intelligent Computing in Medical Sciences \& Image Processing}, vol.~2,
  no.~1, pp. 1--13, 2008.

\bibitem{Malagutti2013}
N.~Malagutti, A.~Dehghani, and R.~A. Kennedy, ``Robust control design for
  automatic regulation of blood pressure,'' \emph{IET Control Theory \&
  Applications}, vol.~7, no.~3, pp. 387--396, 2013.

\bibitem{Malagutti2014}
N.~Malagutti, ``Particle filter-based robust adaptive control for closed-loop
  administration of sodium nitroprusside,'' \emph{Journal of Computational
  Surgery}, vol.~1, no.~1, pp. 1--19, 2014.

\bibitem{Cui2019}
X.~Cui, Z.~He, E.~Li, A.~Cheng, M.~Luo, and Y.~Guo, ``{State-of-charge
  estimation of power lithium-ion batteries based on an embedded micro control
  unit using a square root cubature Kalman filter at various ambient
  temperatures},'' \emph{International Journal of Energy Research}, vol.~43,
  no.~8, pp. 3561--3577, 2019.

\bibitem{luspay2016adaptive}
T.~Luspay and K.~M. Grigoriadis, ``Adaptive parameter estimation of blood
  pressure dynamics subject to vasoactive drug infusion,'' \emph{IEEE
  Transactions on Control Systems Technology}, vol.~24, no.~3, pp. 779--787,
  2015.

\bibitem{loehr2014advanced}
N.~Loehr, \emph{Advanced linear algebra}.\hskip 1em plus 0.5em minus
  0.4em\relax New York, NY, USA: Chapman and Hall/CRC, 2014.

\bibitem{jia2013high}
B.~Jia, M.~Xin, and Y.~Cheng, ``{High-degree cubature Kalman filter},''
  \emph{Automatica}, vol.~49, no.~2, pp. 510--518, 2013.

\bibitem{liu2014adaptive}
Y.~Liu, K.~Dong, H.~Wang, J.~Liu, Y.~He, and L.~Pan, ``{Adaptive Gaussian sum
  squared-root cubature Kalman filter with split-merge scheme for state
  estimation},'' \emph{Chinese Journal of Aeronautics}, vol.~27, no.~5, pp.
  1242--1250, 2014.

\bibitem{hanlon2000multiple}
P.~D. Hanlon and P.~S. Maybeck, ``{Multiple-model adaptive estimation using a
  residual correlation Kalman filter bank},'' \emph{IEEE Transactions on
  Aerospace and Electronic Systems}, vol.~36, no.~2, pp. 393--406, 2000.

\bibitem{sandu2016reinforcement}
C.~Sandu and D.~Popescu, ``Reinforcement learning for the control of blood
  pressure in post cardiac surgery patients,'' \emph{U. P. B. Sci. Bull.,
  Series C}, vol.~78, no.~1, pp. 139--150, 2016.

\bibitem{isaka1993control}
S.~Isaka and A.~V. Sebald, ``Control strategies for arterial blood pressure
  regulation,'' \emph{IEEE Transactions on Biomedical Engineering}, vol.~40,
  no.~4, pp. 353--363, 1993.

\bibitem{rao2003experimental}
R.~R. Rao, B.~Aufderheide, and B.~W. Bequette, ``Experimental studies on
  multiple-model predictive control for automated regulation of hemodynamic
  variables,'' \emph{IEEE Transactions on Biomedical Engineering}, vol.~50,
  no.~3, pp. 277--288, 2003.

\end{thebibliography}
\end{document}